# Wave and Quantum Properties of Peptide Strings: Defining a Helix in Spacetime

**Razvan Tudor Radulescu**

Molecular Concepts Research (MCR), Muenster, Germany

E-mail: ratura@gmx.net

**Mottos:**  *"It is not the outer form that is real, but the essence of things."*
(Constantin Brâncuși)

*"Atoms are not objects, but components of observation situations."*
(Werner Heisenberg)

*"Quantum mechanics will be expanded by biological terms."*
(Niels Bohr)

**Keywords**:   Quantum, Biology, Physics, Cell, Subcellular, Peptide Strings, Particle, Wave, Duality, Periodicity, Resonance, Coherence, Helix, Spacetime, Binary, Information, Cybernetics

**ABSTRACT**

Previous studies have described the concept of peptide strings in qualitative terms and illustrated it by applying its corrolaries in order to elucidate basic questions in oncology and rheumatology. The present investigation is the first to quantify these potential sub- and transcellular phenomena. Accordingly, the propagation of peptide strings is proposed here to occur by way of waves that in turn are subject to the energy equation established by Planck. As a result of these insights, widespread future applications can now be envisaged for peptide strings both in molecular medicine and quantum optics.

Over the past few years, the new interdisciplinary research fields of particle biology (1) and peptide strings (2-4) have been outlined. Moreover, the peptide strings concept was applied to develop novel views in the understanding and treatment of cancer (3,5), thereby expanding the epigenetic dimension of this disease (6), as well as of rheumatoid arthritis (7,8). It is conceivable that these qualitative descriptions primarily involving precisely defined elements of sub/transcellular protein dynamics, bioinformatics and peptide interactions might suffice for future translation into clinical practice. However, additional *quantitative* considerations could be crucial for successfully accomplishing such task and, furthermore, enlarge the scope of peptide strings beyond medicine. Therefore, an initial approximation towards this goal will be attempted here.

　　　Among the key premises of the present quantitative definition of peptide strings is the fact that particles such as electrons have a dual nature in that they may also be perceived as waves (9).

　　　Along similar lines, peptides represent, on the one hand, informational entities and particles in the sense of particle biology (1) or, as I would also coin them, biological quanta. In support of this view is the fact that single amino acids (e.g. glycine) are known to act, for instance, as neurotransmitters in synaptic clefts, but not across cells whereas protein molecules may translocate from one subcellular compartment to another and thereby influence cell fate by virtue of internal and/or external peptide signatures.

　　　On the other hand, the peptides´ potential formation of peptide strings across cells follows a *periodic* character which ensures that a certain, e.g. oncogenic vs. anti-oncogenic (3-6) or arthritogenic vs. anti-arthritogenic (7), peptide information is propagated sub- and transcellularly. Consequently, I propose that it is this *periodicity* that could represent the essence of a wave-like property of distinct peptide sequences resonating over large cellular distances and thus engendering a quantum state of coherence for a given period of time (Fig. 1).

Intrinsic to such state would be a certain energy value amenable to calculation through Planck´s equation (10), i.e. $\varepsilon = h\nu$ or, respectively, $E = h f$ where $E$ is the energy, $h$ represents Planck´s constant and $f$ designates the frequency, the latter of which is the ratio between the velocity $v$ and the wavelength $\lambda$.

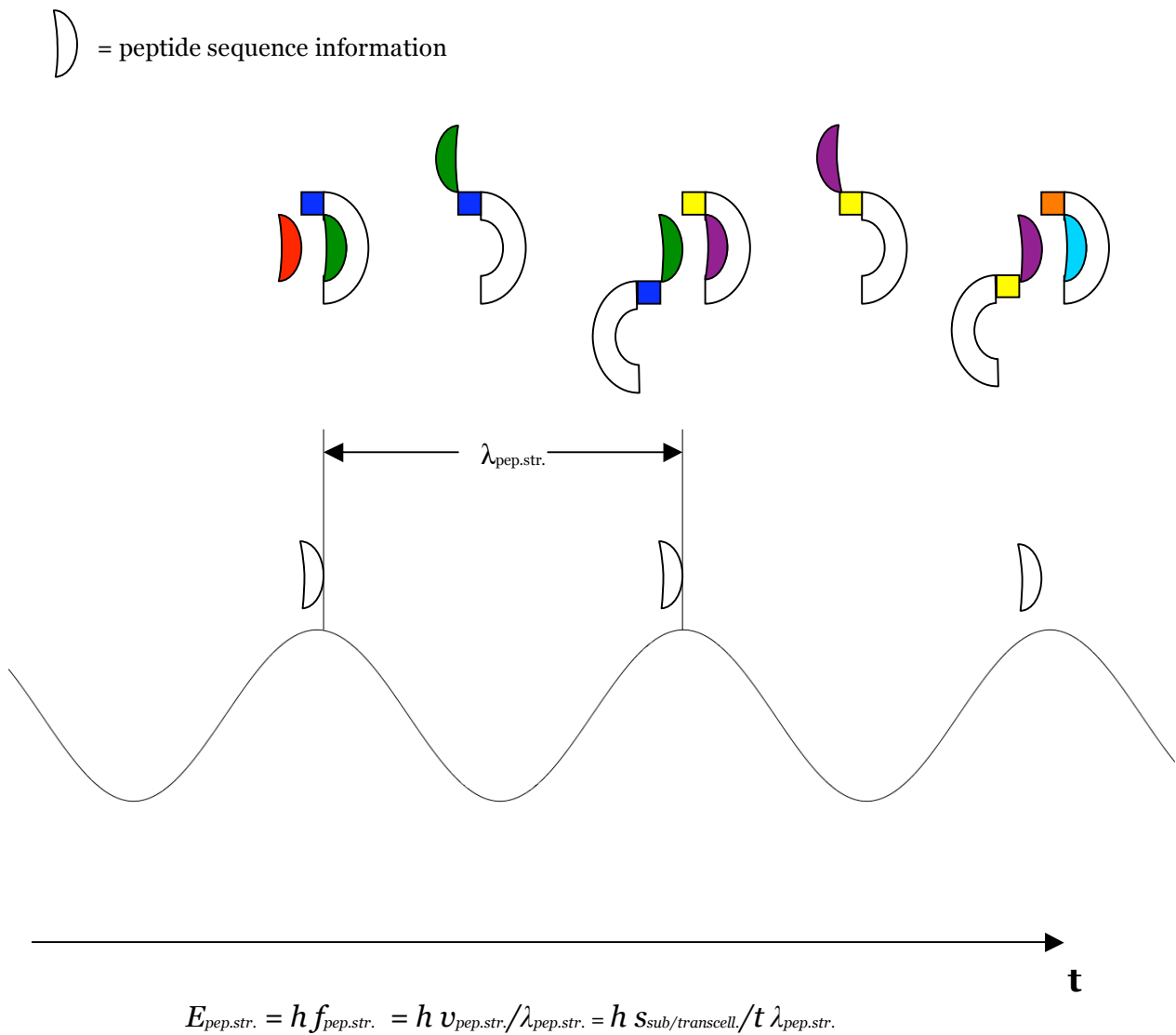

$$E_{pep.str.} = h f_{pep.str.} = h\, v_{pep.str.}/\lambda_{pep.str.} = h\, s_{sub/transcell.}/t\, \lambda_{pep.str.}$$

**Fig. 1**    Schematic representation illustrating the periodic propagation of a given peptide sequence information by means of a peptide string wave, as initiated by a peptide or peptide-like substance (cf. object in red color) and further disseminated by its (protein-embedded) peptide replicas (cf. objects in green and purple). The peptide string wavelength $\lambda_{pep.str.}$ is indicated. $E_{pep.str.}$= energy of peptide string, $h$= Planck´s constant, $f_{pep.str.}$ = frequency of peptide string wave, $v_{pep.str.}$ = velocity of peptide string propagation, $s_{sub/transcell.}$= sub/transcellular distance covered by peptide string wave, t= period of time during which the given peptide string is operative.

The wavelength $\lambda$ of such peptide string could in turn be calculated by applying de Broglie´s equation (9): $\lambda = h/p$ whereby the impulse $p$ is the product of the mass $m$ and its velocity $v$. It follows therefrom that the peptide string wavelength $\lambda_{pep.str.}$ may be determined as follows: $\lambda_{pep.str.} = h/m\,v$. More specifically, $m$ would be here the mass of a given peptide capable of being propagated by means of a peptide string wave and $v$ the latter´s sub/transcellular velocity. Based on a van der Waals distance of 3 Å between the contact residues of a distinct pair of interacting proteins and a duration of 1 ns for a protein´s conformational change (11) resulting from such interaction, I derive a (maximum) peptide string velocity of 0.3 m/s for a peptide with a mass of 1000 daltons which, according to above formula, would yield a (minimum) wavelength of approximately 10 Å, i.e. a value that is tenfold higher than that of the electron wavelength (9).

Interestingly, this anticipated wavelength is close to the diameter range for proteins (11) which reflects the equally allosteric nature of peptide strings whereby the stimulus information is essentially transmitted through protein domains to yield a (peptide) replica of the stimulus that is then propagated in spacetime. Thereby, the energy transfer inherent to such process could conceivably be measured, e.g. in a manner analogous to previously described methods (12,13).

Given further that interactions between proteins involve changes in the electron structures of these molecules (14) which predictably result in modifications of their electromagnetic fields associated with the emission and/or absorption of photons, it is likely that light may be emitted and/or absorbed during the propagation of peptide strings.

Consequently, peptide strings might be used not only for therapeutic purposes (5-8), but also in photonics where they could fundamentally extend the potential of peptides for transmitting photocurrent and acting as photodiodes (15).

Considering previous studies (3,4,7) and the present investigation together, I postulate the following 3 principles or laws of peptide strings:

1) **Upon specific stimulation, a self-complementary peptide/protein (with potential for localization in 2 or more distinct subcellular compartments) relays the stimulus information through a stimulus-like region of its own that binds another self-complementary peptide/protein structurally both similar and complementary to the former one and so forth**. This principle is paralleled by considerations in classical quantum mechanics according to which "The resonator reacted only to those rays which it also emitted, and was not in the slightest bit sensitive to the adjacent spectral regions" (16). Moreover, it could be used for the future development of a LASER-like *peptide information amplification by stimulated expansion of resonance* (PIASER);

2) **Peptide strings are (biological) information and its amplification in one entity**, thus resembling the (physicochemical) bifunctionality of electrons during oxidative phosphorylation. As such, peptide strings possess cybernetic features which, due to their utmost simplicity, surpass even those of the nucleocrine pathway that *per se* provides for information only (17-19);

3) **The higher the concentration of a peptide with the capacity to develop a peptide string wave** or, respectively, that of the protein harboring such peptide, the smaller the peptide string wavelength or, conversely, the higher its frequency and hence also **the higher its energy**. The *rate-limiting factor* for such energy increment would then be the peptide concentration at which (*self*)*aggregation* occurs.

Accordingly, peptide string waves could also be regarded as a helix in spacetime whereby the term "helix" denominates a configuration resulting from repeating the "operation that converts an object into an equivalent object anywhere in space" which in turn is a "translation along an axis coupled with a rotation around the axis", as previously defined (20) and, moreover, paralleling previous wave-helix relationships (21,22).

Through additional explorations, it is likely that such spacetime helices may turn out to be a leverage for overcoming genetic deficiencies, constitute fundamental elements for neural processes and morphogenetic fields, expand binary cybernetics as well as, along with their non-peptide equivalents, contribute to the as yet largely enigmatic (anti-gravitational) dark energy at an astrophysical scale. Essentially, peptides with the potential for peptide strings might be the long-sought link between quantum mechanics and biology, thereby providing a molecular basis for phenomena such as biological coherence (23) and thus ultimately also the emergence of life. In the light of these broad implications, it should be worth further determining the intriguing details of this incipient peptide field.